*Article*

# Employment in Tourism Industries: Are there Subsectors with a Potentially Higher Level of Income?

Pablo Dorta-González [1,*] and Sara M. González-Betancor [2]

1 Institute of Tourism and Sustainable Economic Development (TIDES), Campus de Tafira, University of Las Palmas de Gran Canaria, 35017 Las Palmas de Gran Canaria, Spain
2 Department of Quantitative Methods in Economics and Management, Campus de Tafira, University of Las Palmas de Gran Canaria, 35017 Las Palmas de Gran Canaria, Spain; sara.gonzalez@ulpgc.es
* Correspondence: pablo.dorta@ulpgc.es

**Abstract:** This work analyzes the tourist sector, the employment generated by the tourism industries, and its relationship with tourism receipts. The hypothesis is that there are tourist subsectors with a potentially higher level of income. The article studies the impact of the distribution of the employed population in the different subsectors of the tourism industry, controlling for the most important economic variables, on the level of income per arrival in 24 OECD countries, using panel data for the period 2008–2018. As its main result, the model indicates that the labor force that increases most the receipts per arrival is the 'travel agencies and other reservation services', followed by the 'sports and recreation industry' labor force, while having a large labor force in the 'food and beverage' or 'cultural industry' operates in the opposite direction.

## 1. Introduction

Data are fundamental for informed decision-making in planning, monitoring and measuring the effectiveness of employment policies (see [1] for a critical review of the literature about tourism employment). Tourism statistics and tourism employment related data are necessary to understand tourism labor markets, promoting employment opportunities, planning adequate job creation policies, and developing the human resources through education and training [2–5].

This work addresses empirical research on employment in tourism industry from a broader scope than the organizational and managerial, as recommended in [1]. As these authors state, this is the way to address the political, social and economic perspectives of the issue.

According to a UNWTO World Tourism Barometer (18 January 2020), international tourist arrivals worldwide grew 4% in 2019 to reach 1.5 billion, based on data reported by destinations around the world. In 2021, after several months of unprecedented disruption by the COVID-19 pandemic, restrictions on travel remain in the majority of global destinations, and tourism is one of the most affected sectors.

Hopefully, the sector is beginning to restart in some areas. However, the tourism industries remain inadequately measured and insufficiently studied in terms of employment. Employment in the tourism industries needs to be measured and described in a more consistent way supported by proper statistical instruments [6]. Enhancing the quality and comparability of tourism employment statistics would significantly improve the monitoring of tourism labor markets and the effective use of qualified labor, ensuring sustainable tourism development.

From the empirical data, it seems that the countries with the highest income per tourist are those that have 'other country-specific tourism industries'. While the countries with the lowest income per tourist are those that have most of their workforce in the 'food and beverage' and 'passenger transport' industries. Thus, it seems there is a relationship between the distribution of the labor force in the tourism industries and the level of income per tourist in the countries.

To validate this hypothesis, we estimate a robust panel model in which the endogenous variable is the income logarithm, and the explanatory variables are the logarithm labor force in each of the industries, controlling for some economic country variables.

We discuss in this paper the model with the best fit, after running the relevant robustness check tests. This model explains around 98% of the variation of the dependent variable, which implies that there is a strong relation between the explanatory variables and the dependent variable. Estimations have been done with the STATA 16 program using panel data for 24 OECD countries over the period 2008 to 2018.

This analysis is of interest, since the distribution of employment can characterize the tourism offer and the income of a tourist destination. Moreover, in a scenario like the current one, after COVID-19 it is possible that many tourist destinations may start from zero in their demand and, consequently, opt to make changes in their tourism policy as a tourist destination.

The structure of the paper is as follows. Section 2 presents a review of the literature on employment in the tourism sector. Next, the methodology used is described in Section 3. The results obtained are presented in Section 4 and, finally, the conclusions of the study are addressed in Section 5.

## 2. Literature Review

Tourism is a source of economic growth and job creation in both advanced and emerging economies [7,8]. Notably, one job in tourism industries generates one and a half jobs elsewhere [6]. The tourism industry also offers employment opportunities to groups that tend to have difficulties accessing the labor market, such as low-skilled workers, young people, women and immigrants [8]. Tourism is a labor-intensive sector that in some cases receives workers with no previous experience or who have difficulties finding another job (see [9]). The accommodation industry has a higher percentage of young workers than other tourism industries, in many cases part-time or seasonally [10].

However, there are different views in relation to the tourism employment precariousness and the theoretical debate on sustainable tourism [11,12]. Some authors argue that it is difficult to speak of precariousness or to speak of 'bad jobs', since it is relative. It depends on factors such as the alternative jobs you have, the sectors, and even the positions [13]. Other authors have a critical view of the employment precariousness in the sector [14].

Analyzing the characteristics of employment in the tourism sector is not an easy task. The tourism industry encompasses a very heterogeneous set of businesses, types of contracts with workers and working conditions. Moreover, the tourism employment is characterized by notable differences between regions and between seasons of the year [13,15–17].

Large tourism enterprises are concentrated in accommodation and transport activities, generating a substantial share of total employment. However, tourism creates opportunities for entrepreneurs and many small and micro enterprises, in the formal but also in the informal sector, because of the varied demand of a wide range of visitors [18].

In the accommodation sector there is on average one worker for each room placed on the market. To this must be added the indirect jobs. Thus, for each direct job in the hotel industry, another three jobs are generated in complementary activities such as travel agencies, tourist guides, taxi drivers, bus drivers, airport employees, laundry, gardening, catering, and shops, among others [17].

## 3. Materials and Methods

To answer the question of whether there is a relationship between the distribution of the labor force in the tourism industries and the level of income per tourist in the countries, this work uses data from OECD (OECD.stat) and UNWTO (unwto.org) from the years 2008 to 2018 (data extracted in July 1, 2021) for 24 countries, namely the following 19 OECD economies: Australia, Canada, Denmark, France, Hungary, Iceland, Israel, Italy, Latvia, Lithuania, Mexico, New Zealand, Norway, Slovak Republic, Slovenia, Spain, Switzerland, Turkey, United Kingdom; and the following five non-OECD economies: Brazil, Kazakhstan, Malta, Romania, South Africa. These countries correspond to those for which there are disaggregated data on employment by tourism industries, as the hypothesis to be checked is that there are tourist subsectors with a higher potential level of income than others.

A set of explanatory variables has been selected, relating to the distribution of the labor force among the different industries in the tourism sector, together with other variables related to macroeconomic characteristics of the countries, available at the time of the data query on both websites. It was decided to analyze the evolution over the last few years, starting in 2008, the year in which the international financial crisis began, and to truncate the data in the most current available year for all those OECD countries', namely 2018. Thus, there are 11 time periods and 24 countries for each cross-section, giving a total of 264 observations.

With this information, and after a first approximation to the data in a descriptive and graphical way, different panel data estimations are carried out in STATA 16, with the following reference specification:

$$Y_{it} = \alpha_i + \beta_1 X_{1it} + \beta_2 X_{2it} + \cdots + \beta_k X_{kit} + \varepsilon_{it} \quad (1)$$

where $Y$ is the dependent variable, $i$ refers to countries, $t$ refers to time periods, $\alpha$ represents the intercept, $\beta$ represents the coefficients of each of the independent variables, $X$ is the set of explanatory and control variables, $k$ is the total amount of independent variables to be included in the analysis, and $\varepsilon$ is the error or disturbance term. The set of variables and their source are in Table 1, their mean values for the whole period are in Table A2, and the descriptive statistics for the panel data are in Table A3. This set of explanatory variables (Table A2) meet our purpose of analysis, although inevitably there may be other variables that are also likely to be correlated with the revenue per tourist and that are being omitted due to lack of data availability.

**Table 1.** List of variables.

| Acronym | Variable | Source |
|---|---|---|
| REC | Receipts per arrival, USD (dependent variable) | UNWTO |
| AS | Number of workers in the accommodation services industry (explanatory variable) | OECD |
| CI | Number of workers in the cultural industry (explanatory variable) | OECD |
| F&B | Number of workers in the food and beverage industry (explanatory variable) | OECD |
| PT | Number of workers in the passenger transport industry (explanatory variable) | OECD |
| S&RI | Number of workers in the sports and recreation industry (explanatory variable) | OECD |
| TA&RS | Number of workers in the travel agencies and other reservation services industry (explanatory variable) | OECD |
| GDP_PC | GDP per capita at current prices and current Purchasing Power Parity (PPP) USD (control variable) | OECD |
| PERC | Percentage of workers employed in the tourism industry over the total number of workers (control variable) | OECD |
| OCDE | Dichotomous variable indicating whether the country belongs to the OECD (control variable) | OECD |

Note: All tourism industries listed in the OECD statistics are used as explanatory variables, with the exception of the so-called "Other country-specific tourism industries". The latter is a kind of catch-all that not all countries use and for which, therefore, there are missing observations for most countries. Source: Prepared by the authors

The variables in Equation (1), except PERC and OCDE, are expressed in logarithms, so that the model can be interpreted in terms of multiplicative changes in the dependent variable in response to marginal changes in the explanatory variables.

We start with an ordinary least squares (OLS) estimation in which the time and space dimensions of the data are not taken into account (pooled regression), and compare this model with a fixed-effects panel model and a random-effects panel model. Once the best specification of these three is chosen, different robustness tests are carried out, until the final specification is reached.

By working with a larger sample size, due to the availability of longitudinal as well as cross-sectional data, the panel data technique is able to analyze changes in the dependence of the explanatory variables and the dynamics of these variables. Panel data analysis estimates the relationship between the dependent and the independent variables, so that the coefficients of the latter indicate their relative contribution to the joint prediction, over the time period, for the individuals for whom information is available over time. Furthermore, panel data analysis allows us to control for those factors that are not observable but that affect the individual heterogeneity of countries, or for those phenomena that are homogeneous to them but that change over time (time effects).

## 4. Results

### 4.1. Descriptive Analysis

The distribution of the labor force in the different industries of the tourist sector of a country provides an approximation of the type of tourist offer in that country. Meanwhile, the comparison between total income derived from the tourist sector and income per arrival, allows us to approximate the volume of tourists arriving in each country, while indicating the average level of expenditure of such tourism. Through the following figures we will analyze in detail the behavior of these variables for the 24 countries in our sample.

The countries with the highest volume of employees in the tourist sector, in gross terms and decreasing order, are Spain (with more than 2.5 million), Brazil, Turkey, and Italy (with more than 1.9 million). At the bottom of this ranking, and in increasing order, are Iceland, Malta, and Lithuania, who do not have more than 50,000 employees in this sector (Figure 1).

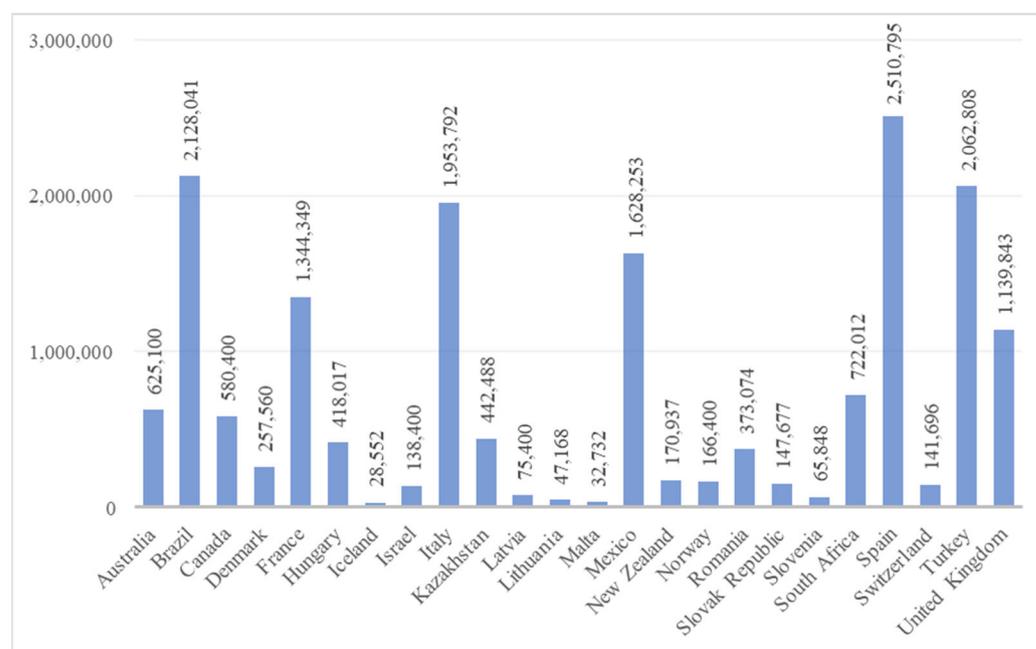

**Figure 1.** Employment in the tourist sector in 2017. Countries in alphabetical order. Source: OECD data for 2017.

With this workforce, the income of the tourist sector in each of these countries is the one shown in Figure 2, both in gross terms (USD billion) and in relative terms (USD per arrival). The countries with the highest level of gross income for international tourism receipts are, in decreasing order: Spain, France, United Kingdom, and Italy. However, the per-arrival income level of these countries is much lower than that of countries such as Australia, New Zealand, and even Israel. The hypothesis is that there are tourist subsectors with a higher potential level of income than others. To check this hypothesis we estimate different panel models to explain the income by the distribution of the labor force in each of the tourist subsectors, controlling for some economic country variables, with data related to the years 2008 to 2018 for the countries in our sample.

The volume of incoming tourists in the former, is much higher than in the latter, so as to generate such a high income. It seems, therefore, that the former are countries with mass tourism with a low level of expenditure as opposed to the latter, which are countries with less tourism but with a higher level of expenditure. It also seems to depend on the remoteness of the tourists' origin and the per capita income of the destination countries, which will influence the tourist stay costs.

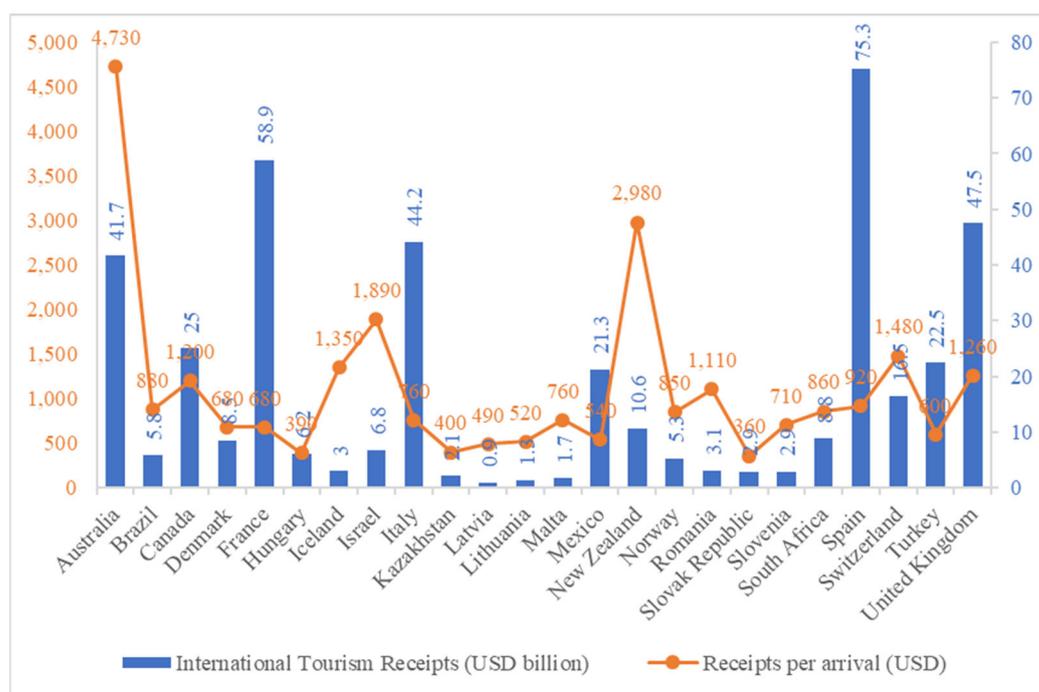

**Figure 2.** Total international tourism receipts, and receipts per arrival in 2017. Countries in alphabetical order. Source: UNWTO data for 2017.

The productivity of the labor force, in terms of annual income per worker in the tourist sector (USD), also differs between countries, as can be deduced from Figure 3. The countries where the labor force in the tourist sector is less profitable are also the countries with a lower GDP, such as Brazil, Kazakhstan or Romania. On the other hand, those countries with a higher return on their labor force coincide with countries with a higher level of GDP, such as Switzerland and Iceland.

The distribution of the labor force among the different industries that make up the tourism sector, according to data from the World Tourism Organization (UNWTO), enables a comparison between countries to see in which industry they are most labor intensive (Figure 4). Countries with a higher number of workers in the tourist sector seem to be mostly intensive in the 'food and beverage' industry. Those countries in the middle of the ranking, in terms of number of tourism workers, distribute their workforce mainly between the 'food and beverage' and 'passenger transport' sectors. Finally, those with a lower volume of workers in the tourist sector seem to have their industry divided almost equally between the 'accommodation services', 'food and beverage', and 'passenger transport' industries. The industries with the lowest relative volume of workers are those of 'travel agencies and other reservation services industry' (in all countries), 'cultural industry' (except in the case of Malta, where it reaches 25%), and 'sports and recreation industry' (except in Kazakhstan and Latvia, where it reaches 17% and 15%, respectively).

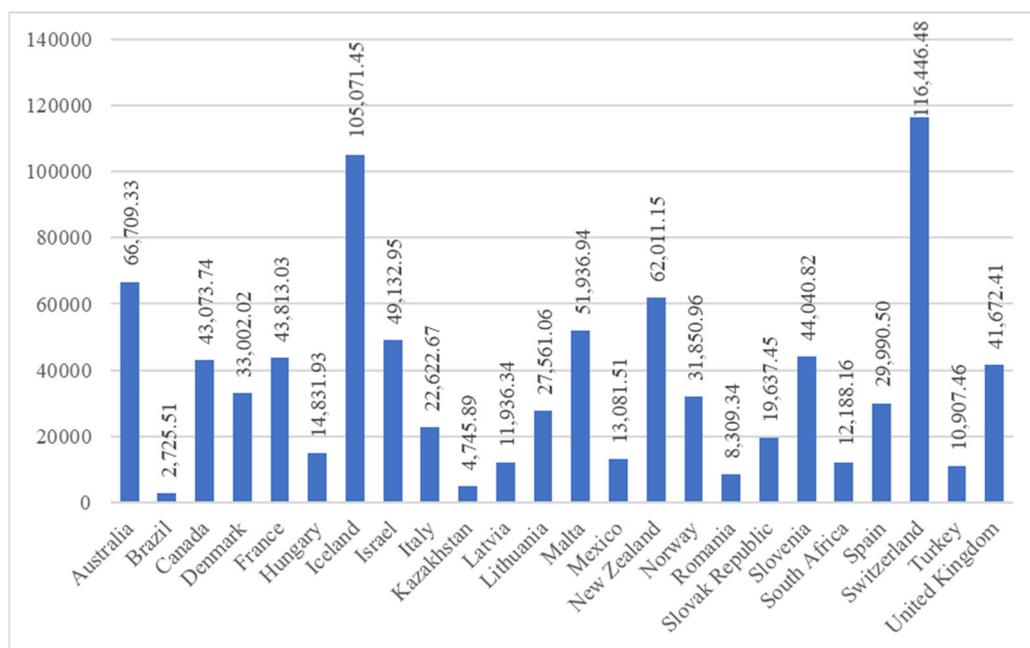

**Figure 3.** International tourism receipts per worker (USD) in 2017. Countries in alphabetical order. Source: UNWTO and OECD data for 2017.

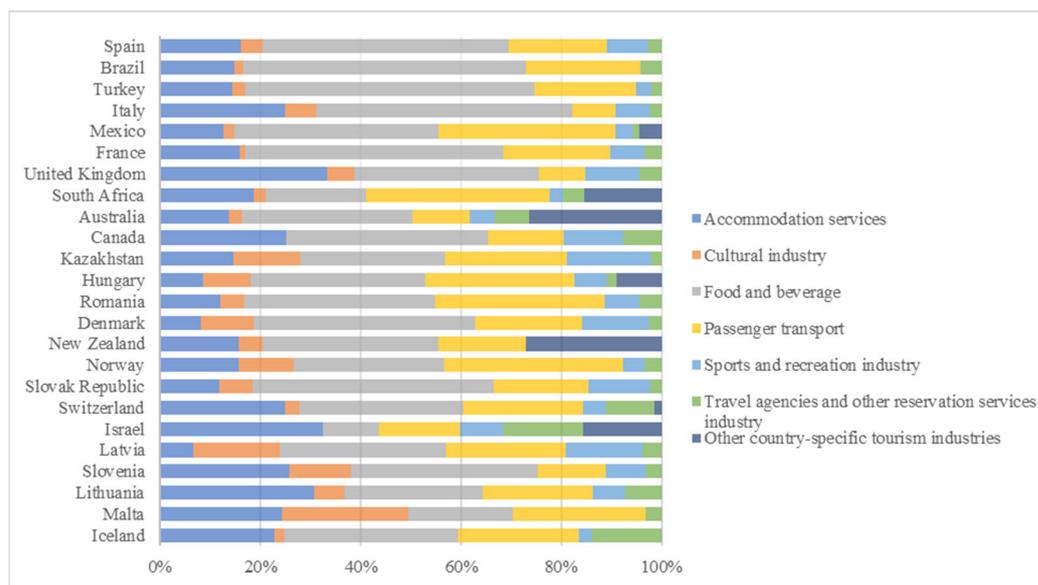

**Figure 4.** Employment percentage distribution in the tourism industries in 2017. Countries ranked in decreasing order by total employment in the tourist sector. Source: UNWTO data for 2017. Note: The percentages for each of the countries and tourism industries can found in the Appendix A.

*4.2. Econometric Analysis*

From the graphs above, it seems that the countries with the highest income per tourist (Australia, New Zealand, and Israel) are those that have their own tourism industries ('other country-specific tourism industries'). While the countries with the lowest income per tourist (Slovak Republic, Hungary, or Kazakhstan) are those that have most of their workforce in the 'food and beverage' and 'passenger transport' industries. Therefore, it is possible that there is some kind of relationship between the distribution of the labor force in the tourism industries and the countries level of income per tourist. In order to check this, and given that countries' data are available for more than one year, we follow a panel data approach—testing different specifications—to explain the variability of the receipts per arrival (in logarithms).

We start with the pooled ordinary least squares (OLS) estimation, in which the time and space dimensions of the data are not taken into account. Since the panel data methodology allows controlling for those country-specific characteristics that are difficult to quantify and are time invariant through the intercept term—$\alpha_i$ in Equation (1), a second estimation has been proposed, in which these unobservable differences across countries are assumed to be constant, i.e., the alpha term is considered fixed, which allows incorporating individual country heterogeneity. Finally, the random effects model considers these unobserved cross-country differences to be random, so that the intercept contains a random component ($\alpha_i = \alpha + u_i$) with a distribution conditional on the regressors (Table 2).

**Table 2.** Panel data estimations of the explanatory variables of tourism receipts (in logarithm) for the countries in the sample.

|  | OLS (Pooled Data) | | Fixed Effects Panel | | Random Effects Panel | |
| --- | --- | --- | --- | --- | --- | --- |
|  | Coef. | Sig. | Coef. | Sig. | Coef. | Sig. |
| L_AS | −0.081 |  | −0.406 | ** | −0.024 |  |
| L_CI | −0.067 |  | −0.062 |  | −0.123 |  |
| L_F&B | −0.462 | *** | −0.955 | *** | −0.061 |  |
| L_PT | −0.045 |  | −0.196 |  | −0.362 | *** |
| L_S&RI | 0.129 |  | 0.038 |  | 0.103 |  |
| L_TA&RS | 0.501 | *** | 0.123 |  | 0.305 | *** |
| L_GDP_PC | −0.244 | ** | 0.030 |  | −0.482 | *** |
| PERC | 0.308 | *** | 0.221 | ** | 0.081 |  |
| OCDE | 0.659 | *** | −0.048 |  | 0.165 |  |
| Constant | 10.015 | *** | 23.012 | *** | 13.986 | *** |
| R-squared | 0.750 | | 0.052 | | 0.350 | |
| R-squared within |  | | 0.311 | | 0.095 | |
| R-squared between |  | | 0.056 | | 0.3884 | |
| F-test/Chi square | 54.351 (F) | | 7.315 (F) | | 35.975 (Chi) | |
| Prob > F | 0.000 | | 0.000 | | 0.000 | |
| Sigma_u |  | | 2.190 | | 0.263 | |
| Sigma_e |  | | 0.174 | | 0.174 | |
| Rho |  | | 0.994 | | 0.696 | |

*** $p < 0.01$, ** $p < 0.05$. Source: Authors' own calculations.

The F-test for fixed effects led us to reject the pooled model against the fixed effects model (F = 25.66; Prob > F = 0.000). The Breusch and Pagan Lagrangian multiplier test for random effects has led us to reject the null hypothesis that the variance of random effects is zero (Chi = 47.66; Prob > Chi = 0.0000), hence the random effects panel data model is better than the pooled regression. Finally, the Hausman test leads us to reject the null hypothesis (Chi = 93.53; Prob > Chi = 0.0000), so the fixed effects model is more consistent and hence the most appropriate, since there continues to be correlation between the error component and the regressors.

We have tried to incorporate time effects, introducing in the fixed effects model a vector of dichotomous variables to check for those events that could have affected all countries equally during the period of time analyzed, thus reducing biases in the estimation (first model of Table 3). The consideration of time effects allows us to include unobservable characteristics that may have influenced tourism receipts in the period analyzed, such as changes in tourist habits after the 2008 financial crisis. However, when testing the joint significance of the time effects, they turn out to be jointly non-significant and therefore do not need to be included in the model (F = 1.70; Prob > F = 0.087).

**Table 3.** Fixed effects panel data estimations for tourism receipts (in logarithm) for the countries in the sample (The full estimations, together with confidence intervals, can be found in the Appendix A (Tables A4–A9)).

|  | Fixed Effects Panel with Time Effects [a] | | Fixed Effects Panel with AR(1) | | Fixed Effects Heteroskedastic Panel with Corrected Standard Errors | |
| --- | --- | --- | --- | --- | --- | --- |
|  | Coef. | Sig. | Coef. | Sig. | Coef. | Sig. |
| L_AS | −0.324 | * | 0.170 |  | −0.019 |  |
| L_CI | 0.014 |  | 0.053 |  | −0.083 | * |
| L_F&B | −0.774 | *** | 0.196 |  | −0.262 | ** |
| L_PT | −0.292 | * | 0.009 |  | −0.08 |  |
| L_S&RI | 0.089 |  | 0.046 |  | 0.126 | *** |
| L_TA&RS | 0.107 |  | 0.048 |  | 0.284 | *** |
| L_GDP_PC | 0.466 |  | 0.079 |  | −0.031 |  |
| PERC | 0.215 | * | −0.541 | *** | 0.256 | *** |
| OCDE | −0.058 |  | 0.006 |  | 0.371 | ** |
| Constant | 15.619 | *** | 1.194 | *** | 10.387 | *** |
| R-squared | 0.090 | | 0.313 | | 0.977 | |
| R-squared within | 0.387 | | 0.420 | | | |
| R-squared between | 0.094 | | 0.351 | | | |
| F-test/Chi square | 4.525 (F) | | 10.288 (F) | | 85.993 (Chi) | |
| Prob > F | 0.000 | | 0.000 | | 0.000 | |
| Sigma_u | 1.828 | | 1.392 | | | |
| Sigma_e | 0.170 | | 0.136 | | | |
| Rho | 0.991 | | 0.991 | | | |

[a] Includes a vector of dichotomous variables from year 2008 (reference) to year 2018. *** $p < 0.01$, ** $p < 0.05$, * $p < 0.1$. Source: Authors' own calculations.

The Woolridge test for autocorrelation in panel data corroborated the existence of autocorrelation (F = 121.223; Prob > F = 0.000). This result is expected since most of the explanatory variables, such as GDP, or even the distribution between tourism sectors, are associated over time. We correct this specification problem by estimating a fixed effects panel with AR(1) (second model of Table 3).

Finally, we check whether the variance of the errors is constant by means of the Modified Wald test. This test for groupwise heteroskedasticity in fixed effect regression model, allowed us to check that the homoskedasticity principle was violated (Chi = 4306.38; Prob > Chi = 0.000), so we estimated the last specification (third model of Table 3), by means of the xtpcse STATA command, which shows the Prais–Winsten regression for heteroskedastic panels with corrected standard errors.

The model explains about 98% of the variation of the dependent variable (the logarithm of receipts per arrival in USD), by means of the independent variables on employment in the seven tourism industries (in logarithms), and three control variables related to the economic characteristics of the country (the logarithm of GDP per capita, the percentage of workers employed in the tourism industry over the country's workers, and the dichotomous variable reflecting OECD membership). This implies that there is a strong relation between the variables in the model and the dependent variable.

A 1% increase in the labor force in the 'cultural' industry (or in the 'food and beverage' industry) implies a 0.08% (or 0.26%) decrease in tourism receipts per arrival, other things being equal. By contrast, if the increase occurs in the 'sports and recreation' industry (or in the 'travel agencies and other reservation services' industry), an increase in tourism receipts per arrival of 0.13% (or 0.28%) is expected, other things being equal. The workforce of all other tourist industries are not statistically significant.

These changes may seem low at first glance. However, it should be noted that in these countries the average number of workers in the tourism sector is 716,723—with the minimum in Iceland and the maximum in Spain (Figure 1)—and that the tourist arrivals in these countries are counted in millions. Therefore, a transfer of thousands of workers from one industry to another can represent a considerable change in the country's total income from tourism.

The control variables introduced in the model allow us to state that receipts per arrival are independent of the country's own wealth, measured by GDP per capita. However, they do depend (strongly) on the share of workers in the tourism sector, relative to the total number of workers in the country (a 1% increase in the share of workers in the tourism sector, relative to the total number of workers in the country, implies a $100 \cdot (e^{0.256} - 1) = 29\%$ increase in arrivals receipts). Finally, OECD countries must share common characteristics that make receipts per arrival higher in OECD countries than in non-OECD countries.

## 5. Conclusions

As mentioned above, the tourist sector is labor intensive. The employment in this sector depends on the tourist offer of each country and the characteristics of the tourism industries. In this paper we analyzed some macroeconomic data for 24 different countries related to the distribution of the tourism labor force among its different industries, as well as to the national income derived from tourism.

The hypothesis is that there are subsectors with a higher potential level of income than others. To check this hypothesis we estimated different specifications of panel models to explain the receipts per arrival with data related to the years 2008 to 2018 for 24 countries. As a main result it can be concluded that the labor force that correlates most positively with the receipts per arrival is the 'travel agencies and other reservation services' labor force, followed by the 'sports and recreation industry' labor force, while having a large labor force in the 'food and beverage' or 'cultural industry' operates in the opposite direction.

This analysis is of interest since the distribution of employment can at least partially characterize the tourism offer and the income of the tourist destination. Furthermore, in a scenario like the current one, after the COVID-19 it is possible that many tourist destinations may start from zero in their demand and, consequently, choose to make changes in their tourism policy as a tourist destination.

Since according to the results of the estimated panel model tourism revenues appear to be directly related to the composition of the country's labor force, it may be in the interest of these countries to try to restructure the labor force to increase the proportion of workers in the tourism sector and, consequently, their revenues in the sector. This could be achieved by bringing more workers, currently unemployed, into the tourism sector or by shifting the country's employees from one sector of activity to the tourism sector. On the other hand, the tourism workforce itself could be restructured by mobilizing employees from less profitable industries to those industries that appear more profitable. However, it should be borne in mind that this mobility—between sectors of activity or within different tourism industries—is neither easy nor free, as it implies that the workers have the skills to develop in the new jobs, which in turn implies costs of soft skills training.

Therefore, a more detailed cost-benefit analysis of this mobility should be carried out for each specific country.

The main limitations of this study are related to the input data. The accessibility of the enlisted countries tourism markets and the purchase power of main tourist target groups. Depending on the countries, the tourism economy is more domestic or international. Moreover, the use of OECD categories by domestic citizens (for instance, food and beverage).

As future research, among the many aspects that affect employment in the tourism sector, technology should be introduced. Although the sector is not highly intensive in technology, there are studies [19] that show that in the most developed countries, and in tourist establishments oriented to the public with a higher purchasing power, there is a progressive substitution effect. Shocks like COVID-19 can accelerate this transformation. The pandemic, as it has done in other sectors, may drive companies in the tourism sector to increase investment in technology and reduce the ratios of employees per customer, or employees per room (in the case of the hotel sector).

**Author Contributions:** Conceptualization, P.D.-G. and S.M.G.-B.; methodology, P.D.-G. and S.M.G.-B.; software, S.M.G.-B.; validation, P.D.-G. and S.M.G.-B.; formal analysis, P.D.-G. and S.M.G.-B.; investigation, P.D.-G. and S.M.G.-B.; resources, P.D.-G. and S.M.G.-B.; data curation, P.D.-G. and S.M.G.-B.; writing—original draft preparation, P.D.-G. and S.M.G.-B.; writing—review and editing, P.D.-G. and S.M.G.-B.; supervision, P.D.-G. All authors have read and agreed to the published version of the manuscript.

**Funding:** This research received no external funding.

**Institutional Review Board Statement:** Not applicable.

**Informed Consent Statement:** Not applicable.

**Data Availability Statement:** This work uses data from OECD (OECD.stat) and UNWTO (unwto.org).

**Conflicts of Interest:** The authors declare no conflict of interest.

## Appendix A

**Table A1.** Percentage distribution by tourism industries of the total employment in the tourist sector in 2017. Countries in alphabetical order. Source: OECD data for 2017.

| Country | AS | CI | F&B | PT | S&RI | TA&RS | OTI |
|---|---|---|---|---|---|---|---|
| Australia | 13.8% | 2.6% | 34.1% | 11.3% | 4.9% | 7.0% | 26.4% |
| Brazil | 14.7% | 1.8% | 56.4% | 22.6% | 0.0% | 4.3% | |
| Canada | 25.2% | 0.0% | 40.2% | 15.0% | 11.9% | 7.6% | |
| Denmark | 8.1% | 10.6% | 44.1% | 21.5% | 13.2% | 2.6% | |
| France | 16.0% | 1.0% | 51.5% | 21.3% | 7.0% | 3.4% | |
| Hungary | 8.6% | 9.4% | 34.9% | 29.7% | 6.5% | 1.9% | 8.9% |
| Iceland | 22.9% | 1.9% | 34.6% | 24.3% | 2.5% | 13.9% | |
| Israel | 32.5% | 0.0% | 11.1% | 16.2% | 8.6% | 16.0% | 15.6% |
| Italy | 25.0% | 6.2% | 50.9% | 8.7% | 6.8% | 2.4% | |
| Kazakhstan | 14.6% | 13.4% | 28.8% | 24.4% | 16.8% | 2.0% | |
| Latvia | 6.5% | 17.1% | 32.9% | 23.6% | 15.3% | 3.7% | 1.0% |
| Lithuania | 30.7% | 6.1% | 27.7% | 21.9% | 6.5% | 7.2% | |
| Malta | 24.3% | 25.2% | 20.8% | 26.5% | 0.0% | 3.1% | |
| Mexico | 12.6% | 2.3% | 40.7% | 35.2% | 3.5% | 1.3% | 4.4% |
| New Zealand | 15.7% | 4.7% | 35.1% | 17.4% | 0.0% | 0.0% | 27.1% |
| Norway | 15.6% | 11.1% | 29.9% | 35.8% | 4.2% | 3.4% | |
| Romania | 12.0% | 4.8% | 38.1% | 33.7% | 7.0% | 4.4% | |
| Slovak Republic | 11.8% | 6.6% | 48.1% | 19.0% | 12.4% | 2.2% | |
| Slovenia | 25.8% | 12.3% | 37.2% | 13.6% | 8.0% | 3.1% | |
| South Africa | 18.8% | 2.3% | 19.9% | 36.6% | 2.7% | 4.3% | 15.3% |
| Spain | 16.0% | 4.4% | 49.2% | 19.6% | 8.2% | 2.6% | |
| Switzerland | 25.0% | 2.7% | 32.7% | 23.9% | 4.6% | 9.6% | 1.4% |
| Turkey | 14.3% | 2.7% | 57.7% | 20.1% | 3.0% | 2.1% | |
| United Kingdom | 33.3% | 5.4% | 36.7% | 9.2% | 10.8% | 4.5% | 0.1% |

AS = Accommodation Services; CI = Cultural Industry; F&B = Food and Beverage; PT = Passenger Transport; S&RI = Sports and Recreation Industry; TA&RS = Travel Agencies and other Reservation Services; OTI = Other country-specific Tourism Industries. Source: Authors' own calculations

**Table A2.** Mean values (2008–2018) of the set of variables for each country.

| Country | REC | AS | CI | F&B | PT | S&RI | TA&RS | GDP_PC | PERC | OCDE |
|---|---|---|---|---|---|---|---|---|---|---|
| Australia | 5546.18 | 81,990.91 | 14,345.45 | 177,009.10 | 62,600.00 | 25,227.27 | 39,136.36 | 46,335.67 | 4.627 | 1.00 |
| Brazil | 1091.27 | 311,997.60 | 35,929.60 | 1,123,011.00 | 450,169.30 | | 94,844.10 | 14,707.81 | 0.686 | 0.00 |
| Canada | 1331.91 | 141,618.20 | | 212,018.20 | 82,254.55 | 66,154.55 | 45,845.45 | 43,917.57 | 0.808 | 1.00 |
| Denmark | 968.09 | 19,562.27 | 26,651.18 | 96,536.36 | 56,273.18 | 37,862.00 | 6304.55 | 47,490.15 | 0.720 | 1.00 |
| France | 789.91 | 212,896.80 | 13,916.36 | 615,889.80 | 275,745.50 | 79,035.09 | 58,444.64 | 39,561.94 | 0.814 | 1.00 |
| Hungary | 620.82 | 36,391.45 | 39,521.09 | 125,954.70 | 84,233.27 | 24,281.91 | 7488.73 | 25,039.63 | 0.904 | 1.00 |
| Iceland | 2359.73 | 4077.73 | 447.00 | 7469.73 | 4210.91 | 685.45 | 1929.82 | 46,815.38 | 2.242 | 1.00 |
| Israel | 2148.09 | 41,471.43 | | 14,370.00 | 16,797.14 | 10,564.29 | 19,447.14 | 33,344.87 | 1.213 | 0.82 |
| Italy | 921.45 | 503,230.00 | 132,709.70 | 101,9512.00 | 178,656.70 | 130,240.70 | 46,295.00 | 37,312.38 | 2.247 | 1.00 |
| Kazakhstan | 457.73 | 457.73 | 7332.64 | 49,855.91 | 95,599.18 | 51,028.64 | 106,643.30 | 65,052.09 | | 0.00 |
| Latvia | 727.82 | 4927.27 | 12,909.09 | 23,236.36 | 19,209.09 | 9781.82 | 2430.00 | 23,019.71 | 0.548 | 0.27 |
| Lithuania | 688.91 | 11,970.18 | 3558.73 | 12,884.55 | 8976.46 | 2638.09 | 3012.09 | 26,484.16 | 0.903 | 0.09 |
| Malta | 936.18 | 7459.36 | 4831.82 | 5553.64 | 3820.82 | 962.44 | 1001.27 | | | 0.00 |
| Mexico | 575.00 | 183,918.80 | 39,076.82 | 612,946.20 | 527,052.50 | 56,133.27 | 17,869.36 | 17,501.78 | 0.375 | 1.00 |
| New Zealand | 3306.91 | 25,302.82 | 7327.09 | 55,020.55 | 9114.82 | | | 35,726.96 | 1.096 | 1.00 |
| Norway | 1030.64 | 23,072.73 | 18,336.36 | 43,945.45 | 50,081.82 | 7154.55 | 5309.09 | 62,582.07 | 0.892 | 1.00 |
| Romania | 236.55 | 44,074.45 | 11,574.09 | 118,293.50 | 125,195.50 | 25,095.45 | 11,485.82 | 22,791.86 | 0.001 | 0.00 |
| Slovak Republic | 1096.55 | 15,569.91 | 9386.18 | 64,349.73 | 25,205.64 | 15,912.45 | 3194.91 | 27,531.09 | 0.646 | 1.00 |
| Slovenia | 1127.91 | 14,949.18 | 7212.46 | 22,648.09 | 5194.64 | 4971.09 | 1885.73 | 31,349.56 | 1.581 | 0.82 |
| South Africa | 1086.64 | 121,794.10 | 19,731.91 | 127,560.90 | 224,410.80 | 18,259.73 | 25,625.64 | 12,351.17 | 0.001 | 0.00 |
| Spain | 1100.18 | 349,488.50 | 98,573.00 | 1,122,579.00 | 270,792.80 | 183,303.70 | 62,509.27 | 34,479.01 | 1.894 | 1.00 |
| Switzerland | 1753.09 | 34,868.45 | 3556.73 | 44,083.91 | 25,929.55 | 5885.27 | 13,750.18 | 61,902.60 | 0.789 | 1.00 |
| Turkey | 885.09 | 293,486.80 | 46,513.40 | 1,161,471.00 | 360,311.20 | 67,107.20 | 48,393.00 | 22,232.43 | 1.102 | 1.00 |
| United Kingdom | 1562.73 | 324,983.90 | 68,648.64 | 407,868.30 | 147,288.20 | 96,512.82 | 89,808.27 | 40,405.60 | 1.064 | 1.00 |

Source: Prepared by the authors.

**Table A3.** Descriptive statistics of the panel.

| Variable | | Mean | Std. Dev. | Min | Max | Observations |
|---|---|---|---|---|---|---|
| REC | overall | 1347.89 | 1120.5 | 151.0 | 6708.0 | N = 264 |
| | between | | 1117.8 | 236.5 | 5546.2 | n = 24 |
| | within | | 231.5 | 328.2 | 2509.7 | T = 11 |
| AS | overall | 102,791.4 | 120,914.8 | 2144.0 | 516,875.0 | N = 245 |
| | between | | 140,746.2 | 4077.7 | 503,230.0 | n = 24 |
| | within | | 13,574.9 | 51,257.5 | 173,807.5 | T bar = 10.2083 |
| CI | overall | 26,198.02 | 27,908.0 | 404.0 | 148,353.0 | N = 227 |
| | between | | 33,394.0 | 447.0 | 132,709.7 | n = 22 |
| | within | | 5855.7 | −10,994.6 | 56,549.4 | T bar = 10.3182 |
| F&B | overall | 261,642.9 | 359,954.2 | 3886.0 | 1,298,528.0 | N = 245 |
| | between | | 404,226.3 | 5553.6 | 1,161,471.0 | n = 24 |
| | within | | 30,529.8 | 147,710.0 | 437,591.9 | T bar = 10.2083 |
| PT | overall | 123,319.9 | 145,554.1 | 2927.0 | 575,459.0 | N = 245 |
| | between | | 149,002.0 | 3820.8 | 527,052.5 | n = 24 |
| | within | | 14,914.1 | 53,883.4 | 180,221.7 | T bar = 10.2083 |
| S&RI | overall | 39,514.79 | 45,806.2 | 298.0 | 208,884.0 | N = 222 |
| | between | | 47,326.8 | 685.5 | 183,303.7 | n = 22 |
| | within | | 9755.3 | −29,786.3 | 87,002.0 | T bar = 10.0909 |
| TA&RS | overall | 25,369.07 | 29,954.0 | 645.0 | 173,697.0 | N = 233 |
| | between | | 28,630.1 | 1001.3 | 94,844.1 | n = 23 |
| | within | | 9637.4 | −10,439.2 | 140,621.4 | T bar = 10.1304 |
| GDP_PC | overall | 34,414.08 | 14,058.38 | 11,483.44 | 71,705.6 | N = 238 |
| | between | | 13,734.03 | 12,351.17 | 62,582.07 | n = 22 |
| | within | | 4003.994 | 25,662.11 | 45,824.5 | T bar = 10.8182 |
| PERC | overall | 1.194634 | 0.9817 | 0.0005 | 5.1216 | N = 231 |
| | between | | 0.9772 | 0.0005 | 4.6274 | n = 22 |
| | within | | 0.2549 | −0.0333 | 2.5441 | T bar = 10.5 |
| OCDE | overall | 0.70833 | 0.455393 | 0 | 1 | N = 264 |
| | between | | 0.436379 | 0 | 1 | n = 24 |
| | within | | 0.155552 | −0.10985 | 1.6174 | T = 11 |

Source: Prepared by the authors.

**Table A4.** Panel data estimations of the explanatory variables of tourism receipts (in logarithm) for the countries in the sample. OLS (pooled data).

| | Coef. | St. Err. | t-Value | *p*-Value | [95% Conf. Interval] | |
|---|---|---|---|---|---|---|
| L_AS | −0.081 | 0.084 | −0.97 | 0.334 | −0.246 | 0.084 |
| L_CI | −0.067 | 0.047 | −1.45 | 0.149 | −0.159 | 0.024 |
| L_F&B | −0.462 | 0.128 | −3.60 | 0.000 | −0.715 | −0.208 |
| L_PT | −0.045 | 0.068 | −0.66 | 0.513 | −0.179 | 0.09 |
| L_S&RI | 0.129 | 0.079 | 1.62 | 0.107 | −0.028 | 0.285 |
| L_TA&RS | 0.501 | 0.075 | 6.64 | 0.000 | 0.352 | 0.65 |
| L_GDP_PC | −0.244 | 0.096 | −2.56 | 0.011 | −0.433 | −0.056 |
| PERC | 0.308 | 0.033 | 9.45 | 0.000 | 0.243 | 0.372 |
| OCDE | 0.659 | 0.105 | 6.29 | 0.000 | 0.452 | 0.866 |
| Constant | 10.015 | 1.071 | 9.35 | 0.000 | 07.9 | 12.13 |

Source: Prepared by the authors.

**Table A5.** Panel data estimations of the explanatory variables of tourism receipts (in logarithm) for the countries in the sample. Fixed effects panel.

| | Coef. | St. Err. | t-Value | *p*-Value | [95% Conf. Interval] | |
|---|---|---|---|---|---|---|
| L_AS | −0.406 | 0.187 | −2.17 | 0.032 | −0.775 | −0.036 |
| L_CI | −0.062 | 0.096 | −0.64 | 0.523 | −0.252 | 0.129 |
| L_F&B | −0.955 | 0.246 | −3.89 | 0.000 | −1.44 | −0.469 |
| L_PT | −0.196 | 0.174 | −1.12 | 0.262 | −0.54 | 0.148 |
| L_S&RI | 0.038 | 0.084 | 0.45 | 0.654 | −0.128 | 0.204 |
| L_TA&RS | 0.123 | 0.106 | 1.16 | 0.247 | −0.086 | 0.332 |
| L_GDP_PC | 0.03 | 0.194 | 0.16 | 0.876 | −0.352 | 0.413 |
| PERC | 0.221 | 0.108 | 2.05 | 0.042 | 0.008 | 0.433 |
| OCDE | −0.048 | 0.089 | −0.53 | 0.596 | −0.224 | 0.129 |
| Constant | 23.012 | 2.423 | 9.50 | 0.000 | 18.222 | 27.801 |

Source: Prepared by the authors.

**Table A6.** Panel data estimations of the explanatory variables of tourism receipts (in logarithm) for the countries in the sample. Random effects panel.

| _rec | Coef. | St. Err. | t-Value | p-Value | [95% Conf. Interval] | |
|---|---|---|---|---|---|---|
| L_AS | −0.024 | 0.131 | −0.18 | 0.854 | −0.282 | 0.233 |
| L_CI | −0.123 | 0.079 | −1.56 | 0.118 | −0.278 | 0.031 |
| L_F&B | −0.061 | 0.176 | −0.35 | 0.729 | −0.405 | 0.283 |
| L_PT | −0.362 | 0.132 | −2.73 | 0.006 | −0.621 | −0.102 |
| L_S&RI | 0.103 | 0.084 | 1.22 | 0.221 | −0.062 | 0.267 |
| L_TA&RS | 0.305 | 0.104 | 2.93 | 0.003 | 0.101 | 0.509 |
| L_GDP_PC | −0.482 | 0.143 | −3.38 | 0.001 | −0.761 | −0.202 |
| PERC | 0.081 | 0.065 | 1.25 | 0.210 | −0.046 | 0.209 |
| OCDE | 0.165 | 0.102 | 1.62 | 0.105 | −0.035 | 0.365 |
| Constant | 13.986 | 1.601 | 8.74 | 0.000 | 10.848 | 17.123 |

Source: Prepared by the authors.

**Table A7.** Panel data estimations of the explanatory variables of tourism receipts (in logarithm) for the countries in the sample. Fixed effects panel with temporary effects.

| l_rec | Coef. | St. Err. | t-Value | p-Value | [95% Conf. Interval] | |
|---|---|---|---|---|---|---|
| L_AS | −0.324 | 0.194 | −1.68 | 0.096 | −0.707 | 0.058 |
| L_CI | 0.014 | 0.11 | 0.12 | 0.902 | −0.204 | 0.231 |
| L_F&B | −0.774 | 0.249 | −3.11 | 0.002 | −1.266 | −0.282 |
| L_PT | −0.292 | 0.175 | −1.67 | 0.098 | −0.639 | 0.055 |
| L_S&RI | 0.089 | 0.088 | 1.02 | 0.310 | −0.084 | 0.263 |
| L_TA&RS | 0.107 | 0.108 | 0.99 | 0.324 | −0.107 | 0.321 |
| L_GDP_PC | 0.466 | 0.295 | 1.58 | 0.116 | −0.117 | 1.048 |
| PERC | 0.215 | 0.112 | 1.92 | 0.056 | −0.006 | 0.437 |
| OCDE | −0.058 | 0.089 | −0.65 | 0.518 | −0.233 | 0.118 |
| 2009 | −0.02 | 0.067 | −0.30 | 0.767 | −0.153 | 0.113 |
| 2010 | −0.042 | 0.067 | −0.63 | 0.530 | −0.176 | 0.091 |
| 2011 | 0.04 | 0.07 | 0.57 | 0.567 | −0.098 | 0.179 |
| 2012 | −0.016 | 0.071 | −0.23 | 0.821 | −0.157 | 0.125 |
| 2013 | −0.098 | 0.074 | −1.32 | 0.188 | −0.245 | 0.049 |
| 2014 | −0.122 | 0.077 | −1.58 | 0.116 | −0.275 | 0.031 |
| 2015 | −0.217 | 0.082 | −2.65 | 0.009 | −0.379 | −0.055 |
| 2016 | −0.227 | 0.099 | −2.29 | 0.023 | −0.423 | −0.031 |
| 2017 | −0.234 | 0.107 | −2.18 | 0.031 | −0.446 | −0.022 |
| 2018 | −0.2 | 0.121 | −1.65 | 0.101 | −0.44 | 0.04 |
| Constant | 15.62 | 4.275 | 3.65 | 0.000 | 7.165 | 24.074 |

Source: Prepared by the authors.

**Table A8.** Panel data estimations of the explanatory variables of tourism receipts (in logarithm) for the countries in the sample. Fixed effects panel with AR(1).

| l_rec | Coef. | St.Err. | t-Value | p-Value | [95% Conf. Interval] | |
|---|---|---|---|---|---|---|
| L_AS | 0.17 | 0.184 | 0.92 | 0.357 | −0.194 | 0.534 |
| L_CI | 0.053 | 0.093 | 0.57 | 0.569 | −0.131 | 0.237 |
| L_F&B | 0.196 | 0.235 | 0.83 | 0.405 | −0.269 | 0.662 |
| L_PT | 0.009 | 0.171 | 0.05 | 0.957 | −0.329 | 0.347 |
| L_S&RI | 0.046 | 0.071 | 0.64 | 0.52 | −0.095 | 0.186 |
| L_TA&RS | 0.048 | 0.091 | 0.52 | 0.603 | −0.133 | 0.228 |
| L_GDP_PC | 0.079 | 0.226 | 0.35 | 0.727 | −0.368 | 0.526 |
| PERC | −0.541 | 0.128 | −4.22 | 0.000 | −0.795 | −0.287 |
| OCDE | 0.006 | 0.091 | 0.07 | 0.946 | −0.173 | 0.185 |
| Constant | 1.194 | 0.302 | 3.96 | 0.000 | 0.597 | 1.791 |

Source: Prepared by the authors.

**Table A9.** Panel data estimations of the explanatory variables of tourism receipts (in logarithm) for the countries in the sample. Fixed effects heteroskedastic panel with corrected standard errors (Prais–Winsten).

| l_rec | Coef. | St.Err. | t-Value | *p*-Value | [95% Conf. Interval] | |
|---|---|---|---|---|---|---|
| L_AS | −0.019 | 0.083 | −0.23 | 0.822 | −0.181 | 0.144 |
| L_CI | −0.083 | 0.044 | −1.86 | 0.063 | −0.17 | 0.005 |
| L_F&B | −0.262 | 0.134 | −1.96 | 0.049 | −0.524 | −0.001 |
| L_PT | −0.08 | 0.094 | −0.85 | 0.394 | −0.263 | 0.104 |
| L_S&RI | 0.126 | 0.04 | 3.14 | 0.002 | 0.047 | 0.205 |
| L_TA&RS | 0.284 | 0.057 | 4.96 | 0.000 | 0.171 | 0.396 |
| L_GDP_PC | −0.031 | 0.11 | −0.28 | 0.782 | −0.246 | 0.185 |
| PERC | 0.256 | 0.058 | 4.40 | 0.000 | 0.142 | 0.369 |
| OCDE | 0.202 | 0.087 | 2.33 | 0.02 | 0.032 | 0.371 |
| Constant | 7.802 | 1.319 | 5.92 | 0.000 | 5.217 | 10.387 |

Source: Prepared by the authors.